\documentclass[times]{TRR}
\pdfoutput=1
\usepackage{moreverb,url}
\usepackage{algorithmic}
\usepackage{graphicx}
\usepackage{textcomp}
\usepackage{wrapfig}
\usepackage{textcomp}
\usepackage{xcolor}
\usepackage{algorithm}
\usepackage{subfigure}
\usepackage{eqparbox} 
\usepackage{amsmath,amssymb,amsfonts}

\newcommand\BibTeX{{\rmfamily B\kern-.05em \textsc{i\kern-.025em b}\kern-.08em
		T\kern-.1667em\lower.7ex\hbox{E}\kern-.125emX}}

\begin{document}
	
	\runninghead{Chen et al.}
	
	\title{Minimizing the Number of Wireless Charging PAD for UAV-Based Wireless Rechargeable Sensor Networks}
	
	\author{Yingjue Chen\affilnum{1}, Yingnan Gu\affilnum{2}, Panfeng Li\affilnum{1} and Feng Lin\affilnum{1}}

	\affiliation{\affilnum {*}This paper has been submitted to International Journal of Distributed Sensor Networks\\
		\affilnum{1}College of Computer Science, Sichuan University, Chengdu, China\\
		\affilnum{2}College of Software Engineering, Sichuan University Chengdu, China}
	
	\corrauth{Feng Lin, College of Computer Science, Sichuan University, Chengdu 610065, China}
	\email{linfeng@scu.edu.cn}

\begin{abstract}
	In wireless rechargeable sensor networks (WRSNs), most of researches address energy scarcity by introducing one or multiple ground mobile vehicles to recharge energy-hungry sensor nodes. The charging efficiency is limited by the moving speed of ground chargers and rough environments, especially in  large-scale scenarios or challenging scenarios such as separate islands. To address the limitations, some researchers consider replacing ground mobile chargers with lightweight unmanned aerial vehicles (UAVs) to support extremely large-scale scenarios, because of the UAV moving at higher speed without geographical limitation. Moreover, multiple automatic landing wireless charging PADs are deployed in the network to recharge UAVs automatically. In this work, we investigate the problem of introducing the minimal number of PADs in UAV-based WRSNs. We propose a novel and adaptive PAD deployment scheme named CDC${\&}$DSC that can adapt to arbitrary locations of the base station, arbitrary geographic distributions of sensor nodes, and arbitrary sizes of network areas. In the proposed scheme, we first obtain an initial PAD deployment solution by clustering nodes in geographic locations. Then, we propose a center shift combining algorithm to optimize this solution by shifting the location of PADs and attempting to merge the adjacent PADs. The simulation results show that compared to existing algorithms, our proposed scheme can use fewer PADs to charge the whole network.
\end{abstract}

\keywords{Wireless rechargeable sensor networks, UAV, PAD deployment, clustering, coverage disk}
	
\maketitle
	
\section{Introduction}
Wireless Rechargeable Sensor Networks (WRSNs) provide a promising approach to prolong the lifetime of sensor networks by introducing mobile chargers to recharge energy-hungry nodes \cite{b1,b2,b3}. In the most of researches, mobile chargers are generally ground vehicles, such as mobile cars and intelligent robots.

However, there are some limitations on using ground vehicle chargers. It is common that sensor nodes may be deployed in complex and cross-distribution geographical conditions such as rivers, land, islands, and rugged mountains. In these challenging scenarios, the complex terrain may hinder the movement of ground vehicles from one node to another. On the other hand, the lightweight Unmanned Aerial Vehicle (UAV) with high speed can adapt to large-scale scenarios or challenging scenarios regardless of geographical constraints \cite{b4,b5}.

A few studies have considered introducing the UAV as a mobile charger in WRSNs \cite{b4,b5,b6,b7}. In large-scale scenarios or challenging scenarios, the UAV works better than ground chargers for charging nodes but it consumes more energy in flight. It is impracticable to directly increase the battery size because of the negative impacts on cost and performance \cite{b8}. Therefore, \cite{b9,b10} developed a Wireless Charging Station (PAD) to recharge the UAV automatically based on some special Wireless Power Transmission (WPT) systems \cite{b11,b12}. By introducing PADs, the UAV can get energy supplement during charging flights in large-scale scenarios or challenging scenarios. \cite{b13} first introduced PADs into UAV-based WRSNs and proposed four heuristic algorithms to minimize the number of PADs. However, these algorithms can only work in the network with uniform node distribution and the central base station(BS). Obviously, they are not practicable and adaptable for typical application scenarios of UAV-based WRSNs.

In this work, inspired by \cite{b13}, we investigate the problem of introducing the minimal number of PADs in a UAV-based WRSN (Minimizing the Number of Deployed PADs, MNDP). Unlike the work in \cite{b13}, our work aims to minimize the number of PADs in scenarios with arbitrary node distribution and arbitrary BS location to exploit the advantages of the UAV. We first define and formulize this problem and then propose a novel PAD deployment scheme Clustering-with-double-constraints and Disks-shift-combining (CDC${\&}$DSC) to address this problem. CDC${\&}$DSC scheme works in two phases. In the first phase, we propose the CDC (Clustering-with-double-constraints) algorithm, to generate an initial solution to our problem. In the second phase, we propose the DSC (Disks-shift-combining) algorithm to optimize the initial solution by shifting the locations of PADs to their nearest neighbors and trying to merge the adjacent PADs. Simulations show that our scheme can adaptively deploy fewer PADs in UAV-based WRSNs than any scheme in \cite{b13}.
\section{Literature Reveiw}
Most of the existing studies in WRSNs focused on using ground vehicles recharging nodes. \cite{b1} designed an algorithm to maximize charging throughput. \cite{b2} focused on the joint optimization of data gathering and energy replenishment. Moreover, \cite{b3} proposed a multi-chargers collaborative charging scheme to improve the shortages of a single charger in large-scale WRSNs. However, the speed limitations and moving obstructive challenges of ground chargers cannot adapt to large-scale scenarios or challenging scenarios.
	
Recent work applies the lightweight UAV in WRSNs to adapt to large-scale scenarios or challenging scenarios to overcome geographical barriers. \cite{b15} developed a specialized WPT system that supports the UAV to transfer power to ground sensor nodes. \cite{b4} proposed a spatial discretization scheme to obtain a UAV charging spot set and maximize the overall charging energy. \cite{b5} considered the UAV hovering energy consumption based on the model by \cite{b14} to minimize the number of hovering points. Considering the shortages of a single UAV, \cite{b7} discussed the multi-UAVs charging problem to minimize the number of UAVs.
	
Although these above schemes have improved the energy efficiency of UAVs during charging tasks, it is still necessary to recharge UAVs  to extend the working time and the working range of UAVs, especially in large-scale or challenging WRSNs \cite{b8}. \cite{b11} designed a lightweight and efficient WPT system to charge UAV and \cite{b12} achieved the larger charging area by multiple extended coils. Further, some researchers developed a small-scale charging station that can be deployed flexibly. \cite{b9} devised an automatic landing pad (PAD) to transfer energy wirelessly through a pair of lightweight induction coils when the UAV is hovering. \cite{b10} introduced two automatic charging platforms with direct contact way and wireless way to adapt to different kinds of UAVs. 
	
Based on the above studies, \cite{b13} introduced PADs into UAV-based WRSNs, so that the UAV with low residual energy can fly to a nearby PAD and replenish its energy. To minimize the number of PADs, \cite{b13} proposed four heuristic algorithms for PAD deployment, namely MSC, TNC, GNC, and DC algorithms. However, these algorithms only consider the central BS and the uniform distribution of nodes, and cannot verify the superiority of UAVs in large-scale scenarios or challenging scenarios.
	
\section{Preliminaries} 
In this section, we first introduce the network model and the UAV consumption model, and then define and formulize the MNDP problem.
\subsection{Network Model}\label{nm}
We consider a large monitoring area with a BS, N static sensor nodes, a UAV as the mobile charger, and several PADs. Let $S = \{ {s_1},{s_2},...,{s_N}\}$ denotes the set of sensor nodes and $P = \{ {p_1},{p_2},...,{p_M}\}$ denotes the set of PADs. Since BS and PADs work similarly when charging the UAV, for convenience of description, they are sometimes collectively referred to as the charging stations in this work. Let ${p_0}$ denotes the BS and ${P^{'}} = \{ {p_0},{p_1},{p_2},...,{p_M}\}$ denotes the set of charging stations. We also use ${p_i} \in P'$ to denote the location of the PAD/BS deployed. To simplify the problem, the energy supply of each charging station (the BS or a PAD) is unlimited.

Each sensor node ${s_i} \in S$ is powered by a rechargeable battery with energy capacity $e$ and deployed statically on a given location $l({s_i})$ that is known by the BS and the UAV. However, the distribution of all nodes deployment is unknown. The BS is also deployed arbitrarily according to the network requirement and it is responsible for data gathering, charging schedule, and serving the UAV.  

\subsection{UAV Consumption Model}\label{um}
The UAV is powered by a rechargeable battery with high capacity ${E_{max}}$. The UAV charges sensor nodes in the point-to-point charging pattern, which means the UAV has to hover over and fully charge only one node. The UAV flies at a constant speed ${V_U}$ between nodes, PADs and the BS. The UAV has to fly to the nearest charging station before its energy runs off. During the charging process, the UAV can only get recharging from the charging stations. The UAV departs from a charging station with full energy to charge sensor nodes. We call the duration between the fully charged UAV leaving a charging station and landing at the next charging station for recharging as one charging flight.

The energy consumption of the UAV comprises of two parts, namely the energy consumed by traveling ${E_{travel}}$ and the energy consumed by charging ${E_{charge}}$. Particularly, the traveling energy consumption can be divided into two parts: one is the energy consumption of flying between moving targets (nodes/BS/PADs) ${E_{mov}}$; the other is the energy consumption of hovering over nodes to perform the energy transferring ${E_{hov}}$. We have the UAV consumed energy ${E_{consume}}$ as follows:
\begin{equation}
	\begin{aligned}\label{eq1}
		{E_{consume}} & = {E_{travel}} + {E_{charge}}\\
		& = {E_{mov}} + {E_{hov}} + {E_{charge}}
	\end{aligned}
\end{equation}

Let’s assume the energy transfer efficiency between the UAV and the charged node is $\eta$. We have
\begin{equation}\label{eq2}
	{E_{charge}} = {E_{rec}}/\eta
\end{equation}
where ${E_{rec}}$ is the total amount of energy that charged nodes need to be replenished.

We calculate the moving power ${P_{mov}}$ and the hovering power ${P_{hov}}$ of the rotary-wing UAV according to the propulsion power formula derived by \cite{b14}. The moving energy ${E_{mov}}$ and the hovering energy ${E_{hov}}$ for one charging flight can be calculated as follows:
\begin{equation}\label{eq3}
	{E_{hov}} = {P_{hov}}{t_h} = ({P_0} + {P_i}){t_h}
\end{equation}
\begin{equation}\label{eq4}
	{E_{mov}} = {P_{mov}}{t_m}
\end{equation}
where ${t_m}$ and ${t_h}$ are the flying time and the hovering time in the current charging flight, respectively. Since $e \ll {E_{max}}$, the time that one sensor node get fully charged can be approximate as a constant $\Delta$. Therefore, we have
\begin{equation}\label{eq5}
	{t_h} = n\Delta
\end{equation}
where ${n}$ is the number of charged nodes in the current charging flight. According to pre-defined ${V_U}$ and ${E_{max}}$, we calculate the maximum flight distance of the UAV ${d_{max}}$ as follows:
\begin{equation}\label{eq6}
	{d_{max}} = \frac{{{E_{max}}}}{{{P_{mov}}}} \times {V_U}
\end{equation}

\section{The Problem of Definition}
Let’s first consider the case without introducing PADs. If no PADs in the network, the UAV only can get charged at the BS. According to the last subsection, the flight radius of the UAV is $\frac{1}{2}{d_{\max }}$. Considering the UAV has to have enough energy to return to the BS after charging one node, the distance  ${d_{cover}}$ between the farthest rechargeable node and the BS has to satisfy the following equation:
\begin{equation}\label{eq7}
	\frac{{\rm{1}}}{{\rm{2}}}{d_{max}} > {d_{cover}} \ge \frac{{\rm{1}}}{{\rm{2}}}\frac{{({E_{max}} - {P_{hov}}\Delta  - e/\eta )}}{{{P_{mov}}}} \times {V_U}
\end{equation}

To be simplified, we can use equation (\ref{eq8}) in our work.
\begin{equation}\label{eq8}
	{d_{cover}}{\rm{ = }}\frac{{\rm{1}}}{{\rm{2}}}\frac{{({E_{max}} - {P_{hov}}\Delta  - e/\eta )}}{{{P_{mov}}}} \times {V_U}
\end{equation}

We call ${d_{cover}}$ as the charging coverage range of the UAV. Obviously, in the case without PADs, the UAV only can charge nodes in a cycle area with the location of BS as the center and ${d_{cover}}$ as the radius. Therefore, to charge the nodes outside this cycle area of BS, we have to introduce PADs to replenish the energy of the UAV. 

In our work, to easily depict, given a node ${s_i}$ and a charging station ${p_j}$, if the distance between ${s_i}$ and ${p_j}$ is less than ${d_{cover}}$, we call ${s_i}$ is covered by ${p_j}$ and denote this relationship as ${s_i} \in C({p_j})$, where $C({p_j})$ is the set of all the nodes covered by ${p_j}$. Let $A({p_j})$ denotes the cycle area ${p_j}$ covered, we also call it as the coverage disk of ${p_j}$. 

Once PADs are deployed, to ensure each node is chargeable, each node needs to be covered by at least one charging station. We call this requirement for PAD deployment as the {\bf coverage constraint}.

On the other hand, since a single UAV is used in our case, to ensure each node is chargeable, the UAV needs to fly between two charging stations. In this way, the UAV can charge the nodes in a new charging coverage disk after charging all the nodes in the current charging coverage disk. This requirement can be represented as the following: Let $dis({p_i},{p_j})$ denotes the distance between ${p_i}$ and ${p_j}$. Given a graph $G = (V,E)$, where $V = P'$ and $E{\rm{ = \{ (}}{p_i},{p_j})|dis({p_i},{p_j}) < {d_{max}}\}$. The $G$ must be a connected graph. We refer to this requirement for PAD deployment as the {\bf connectivity constraint}.

Based on the above analysis, we can define the problem of minimizing the number of deployed PADs in the UAV-based WRSN below:

\newtheorem{definition}{Definition}

\begin{definition}
	The problem of minimizing the number of deployed PADs
	
	Given a monitoring area $\Omega$, a BS ${p_0}$ and a set of nodes $S = \{ {s_1},{s_2},...,{s_N}\}$, a UAV, the objective of the MNDP problem is to find some locations to deploy PADs $P = \{ {p_1},{p_2},...,{p_M}\}$, which minimizes the number of PADs ${M}$, under the coverage constraint and the connectivity constraint. i.e.
	\begin{equation}\label{eq9}
		Min{\rm{ }}\quad{M}
	\end{equation}
	\begin{equation}\label{eq10}
		dis\left( {{s_i},{p_j}} \right) \le {d_{cover}},\forall {s_i} \in S,\exists {p_j} \in P'
	\end{equation}
	\begin{equation}\label{eq11}
		{\rho _{i,j}} = \left\{ {\begin{array}{*{20}{l}}
				{1,{\rm{ if\;}}dis\left( {{p_i},{p_j}} \right) \le {d_{max}}}\\
				{{\rm{0, otherwise }}}
		\end{array}} \right.
	\end{equation}
	\begin{small}
		\begin{equation}\label{eq12}
			\begin{split}	
				\sum\limits_{|\Omega |} {\left( {{\rho _i}{,_{\Omega (1)}} \times \left( {\prod\limits_{k = 1}^{|\Omega | - 1} {{\rho _{\Omega (k),\Omega (k + 1)}}} } \right) \times {\rho _{\Omega (|\Omega |),j}}} \right)} \ge 1,\\
				\forall {p_i},{p_j} \in P'
			\end{split}
		\end{equation}
	\end{small}
	
	Formula (\ref{eq10}) ensures the coverage for all nodes with PADs and the coverage radius is different from \cite{b13}. Formula (\ref{eq11}) and (\ref{eq12}) demonstrate the connectivity for PADs, in (\ref{eq12}) $\Omega$ is a permutation of a subset of $P$ which means the UAV can fly from $p_i$ to $p_j$ through at least one path combined by other PADs. The problem is a NP-Hard problem deduced by \cite{b13}.
\end{definition}
			
\section{Clustering-with-double-constraints ${\&}$ Disks-shift-combine Algorithm (CDC${\&}$DSC)}
\begin{figure*}[tp]
	\centering
	\subfigure[Initial]{
		\begin{minipage}[t]{0.25\linewidth}
			\centering
			\includegraphics[width=\textwidth]{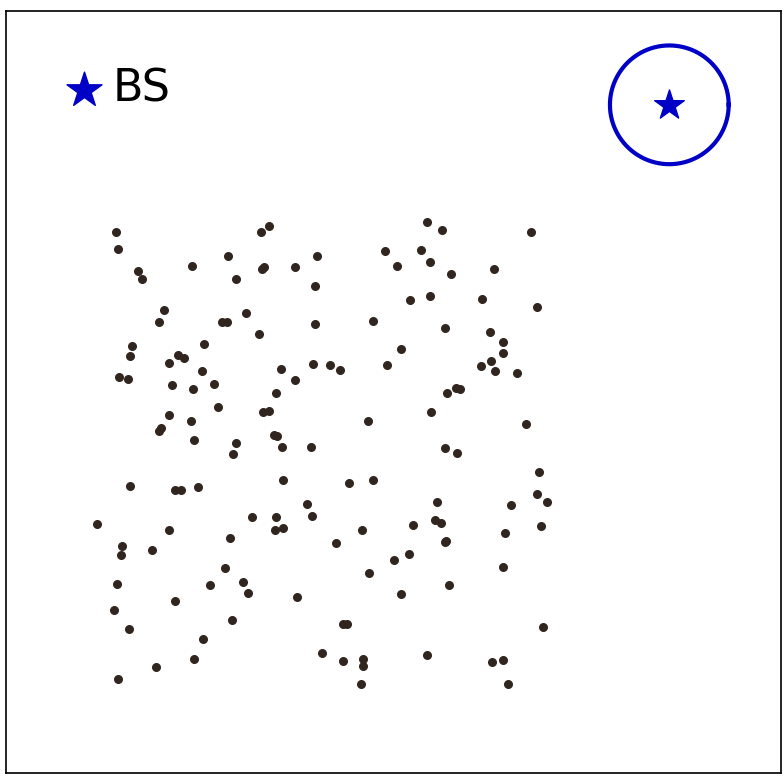}
		\end{minipage}
		\label{fig.1a}
	}%
	\subfigure[After clustering]{
		\begin{minipage}[t]{0.25\linewidth}
			\centering
			\includegraphics[width=\textwidth]{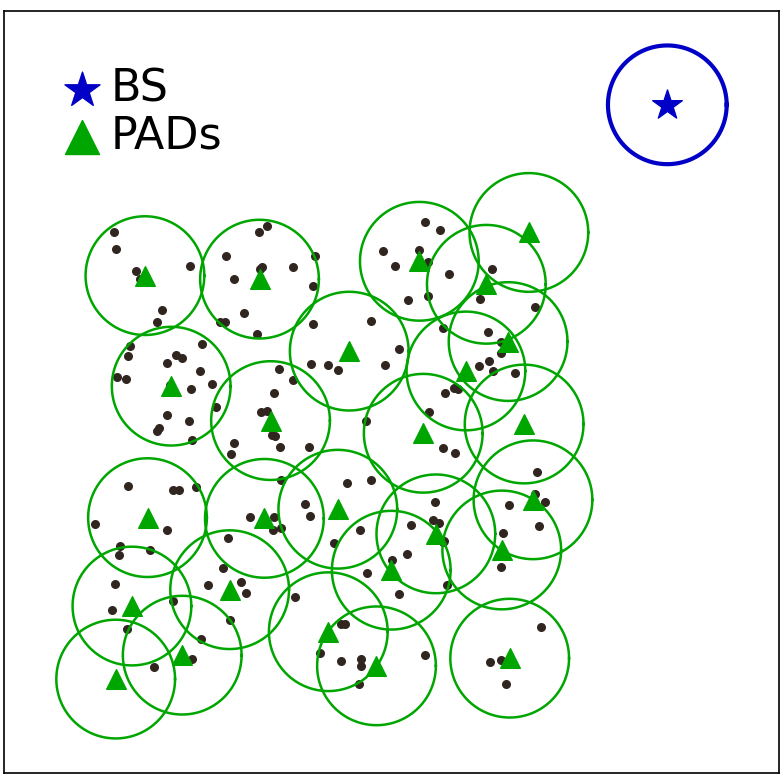}
		\end{minipage}
		\label{fig.1b}
	}%
	\subfigure[After ensuring coverage constraint]{
		\begin{minipage}[t]{0.25\linewidth}
			\centering
			\includegraphics[width=\textwidth]{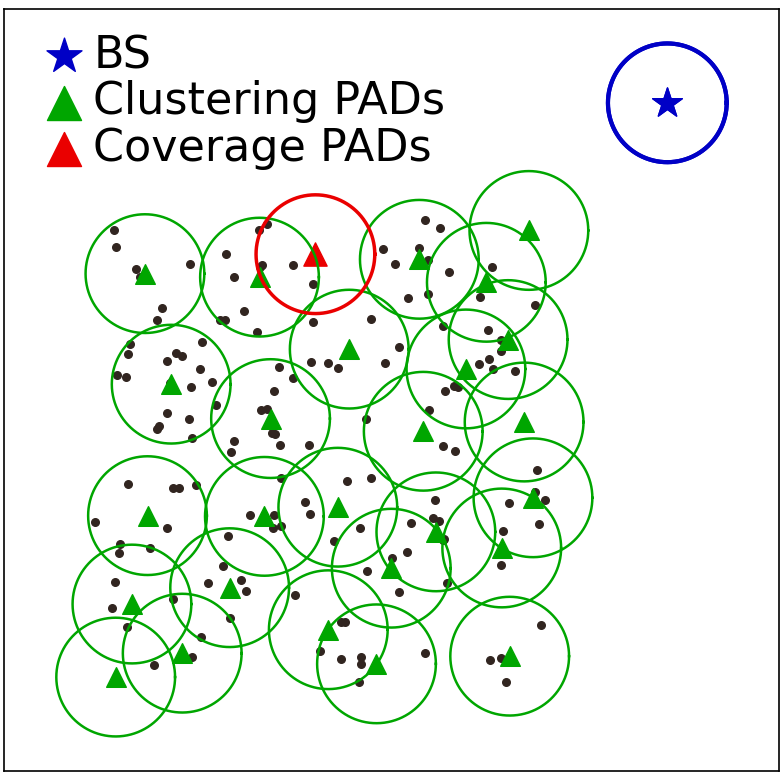}
		\end{minipage}
		\label{fig.1c}
	}%
	\subfigure[After ensuring connectivity constraint]{
		\begin{minipage}[t]{0.25\linewidth}
			\centering
			\includegraphics[width=\textwidth]{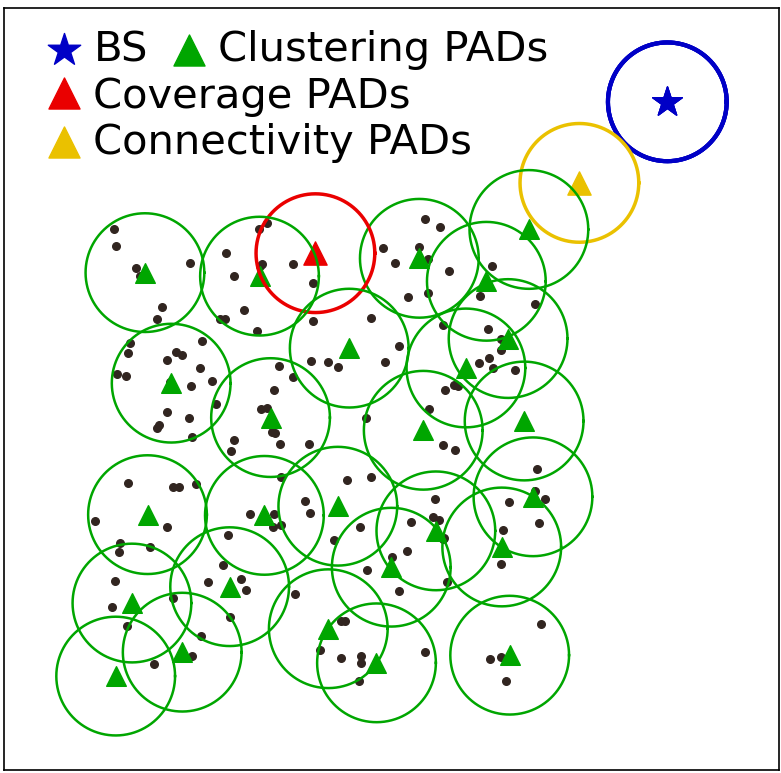}
		\end{minipage}
		\label{fig.1d}
	}
	\centering
	\caption{The working process of the CDC algorithm.~\ref{fig.1a} depicts the location of the nodes and BS in a network area, and~\ref{fig.1b} shows the PADs deployment after clustering the data in~\ref{fig.1a} and generating the coverage disk for each PAD, while~\ref{fig.1c} and~\ref{fig.1d} illuminate the results to meet the coverage constraint and the connectivity constraint, respectively.}
	\label{fig.1}
\end{figure*}
We propose a PAD placement scheme, named CDC\&DSC (Clustering-with-double-constraints \& Disks-shift-combining), to address the MNDP problem, which can automatically adapt to network scenarios with arbitrary node distribution and arbitrary BS location. The CDC\&DSC scheme works in two phases and we propose two algorithms, CDC (Clustering-with-double-constraints) algorithm and DSC (Disks-shift-combining) algorithm, to accomplish the task of each phase separately. The task of phase 1 is to generate the initial solution of the MNDP problem, which we achieve by using the CDC algorithm. In the CDC algorithm, we first cluster all nodes based on geographic locations, then deploy a PAD at the center of each cluster, and then obtain an initial solution to our problem by adding new PADs to satisfy the coverage constraint and the connectivity constraint. The task of phase 2 is to optimize the initial solution from phase 1, which we complete by using the DSC algorithm. In the DSC algorithm, we decrease the number of PADs by combing the PADs based on the principle of triangle circumcircle after trying to shift each PAD to its nearest neighbor. 
In the rest of this section, we depict the details of the two algorithms one by one.

\subsection{CDC Algorithm}\label{CDC}
The goal of the CDC algorithm is to generate an initial solution to the MNDP problem. Essentially, the output of CDC should be a set of PADs with their coverage disks. 
\renewcommand{\algorithmiccomment}[1]{\hfill\eqparbox{COMMENT}{\ #1}}
\begin{algorithm}  
	\caption{CDC Algorithm}  
	\label{CDC algorithm}  
	\hspace*{0.02in}{\bf Input:}
	A sensor node set $S = \{ {s_1},{s_2},...,{s_N}\}$, a BS\\
	\hspace*{0.02in}{\bf Output:} 
	An initial PAD set ${P}$  
	\begin{algorithmic}[1] 
		\STATE {${P_c} \leftarrow \emptyset$}
		\STATE {$I = \{ {s_i}|{d_i} > \sum\limits_{{s_j} \in S'} {{d_j}} /\left| {S'} \right|\}, K = \left| I \right|*\alpha$}
		\STATE {Calculate $P$ by K-means algorithm with ${S^{'}}$ in which $K = \left| I \right|*\alpha $, the initial points}
		\STATE {Update $C({p_i})$ for each ${p_i} \in P$}
		\STATE {$S' = S - \bigcup\limits_{{p_i} \in P} {C({p_i})} $}
		\WHILE [/*Check coverage constraint*/]{$S' \ne \emptyset $} 
		\STATE Deploy a PAD ${p^{'}}$ on $l({s_i})$ with the largest ${d_i}$
		\STATE {$S' = S - C({p^{'}}),P \leftarrow {\rm{\{ }}{p^{'}}{\rm{\} }}$}
		\ENDWHILE
		\STATE {$G = (V,E),V \leftarrow {\rm{\{ }}BS{\rm{\} }},E \leftarrow \emptyset $}
		\WHILE [/*Check connectivity constraint*/]{$P \ne \emptyset$}
		\STATE {Select ${p_i} \in P$ with the minimal $d({p_i},{p_j}),{p_j} \in V$}
		\IF {$d({p_i},{p_j}) > {d_{max}}$}
		\STATE {Deploy a PAD ${p^{'}}$ on the line connecting ${p_i}$ and ${p_j}$}
		\STATE {$V \leftarrow \{ {p^{'}}\} ,E \leftarrow \{ ({p^{'}},{p_j})\}$}
		\ELSE
		\STATE {$V \leftarrow \{ {p_i}\} ,E \leftarrow \{ ({p_i},{p_j})\} ,P - \{ {p_i}\}$}
		\ENDIF
		\ENDWHILE
		\STATE {$P = V$}
		\label{a1}
	\end{algorithmic} 
\end{algorithm}

Considering the MNDP problem, we must cover all the nodes with as few PADs as possible. Intuitively, the ideal situation is that each node is covered by one and only one PAD. Furthermore, considering the coverage constraint, the geographical locations of nodes covered by the same PAD should be close. Based on the above considerations, we first use a clustering algorithm to cluster the nodes and deploy a PAD in the center of each cluster. Then we can obtain an initial solution to the MNDP problem by adding some PADs to meet the coverage constraint and the connectivity constraint.
\begin{figure*}[tp]
	\centering
	\subfigure[after deleting redundant]{
		\begin{minipage}[t]{0.3\linewidth}
			\centering
			\includegraphics[width=\textwidth]{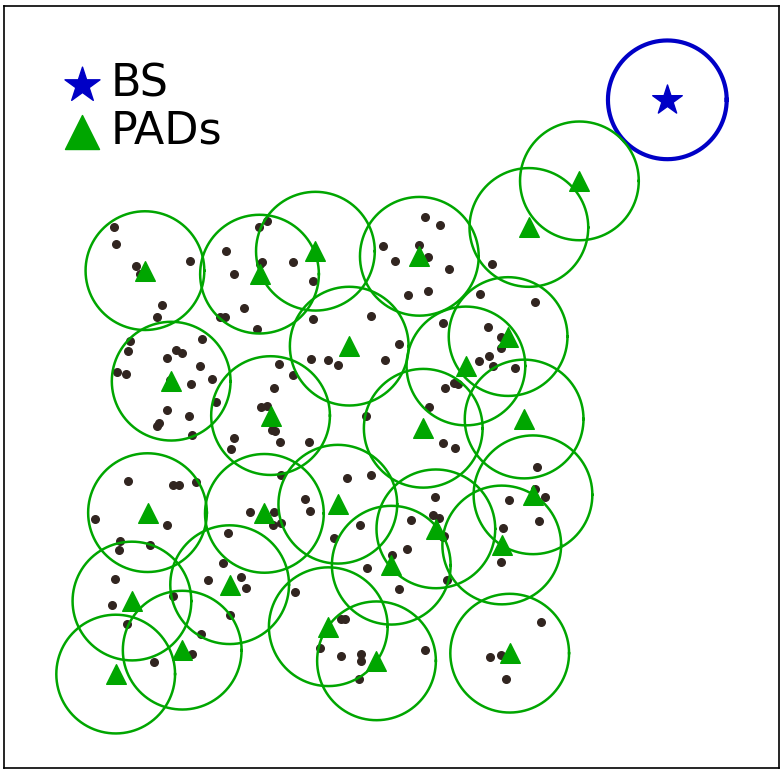}
		\end{minipage}
		\label{fig.2a}
	}
	\subfigure[after shifting]{
		\begin{minipage}[t]{0.3\linewidth}
			\centering
			\includegraphics[width=\textwidth]{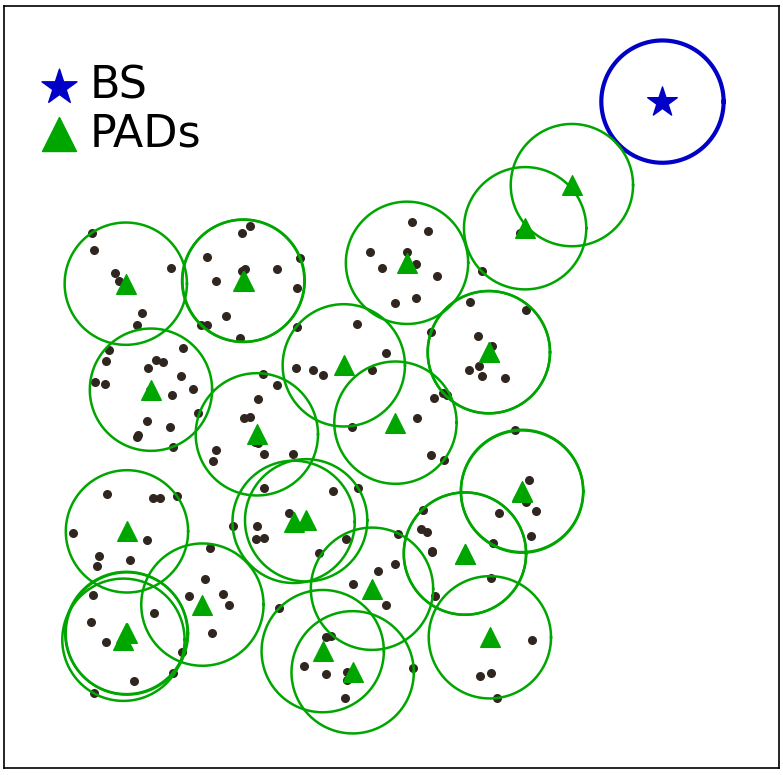}
		\end{minipage}
		\label{fig.2b}
	}
	\subfigure[after combining]{
		\begin{minipage}[t]{0.3\linewidth}
			\centering
			\includegraphics[width=\textwidth]{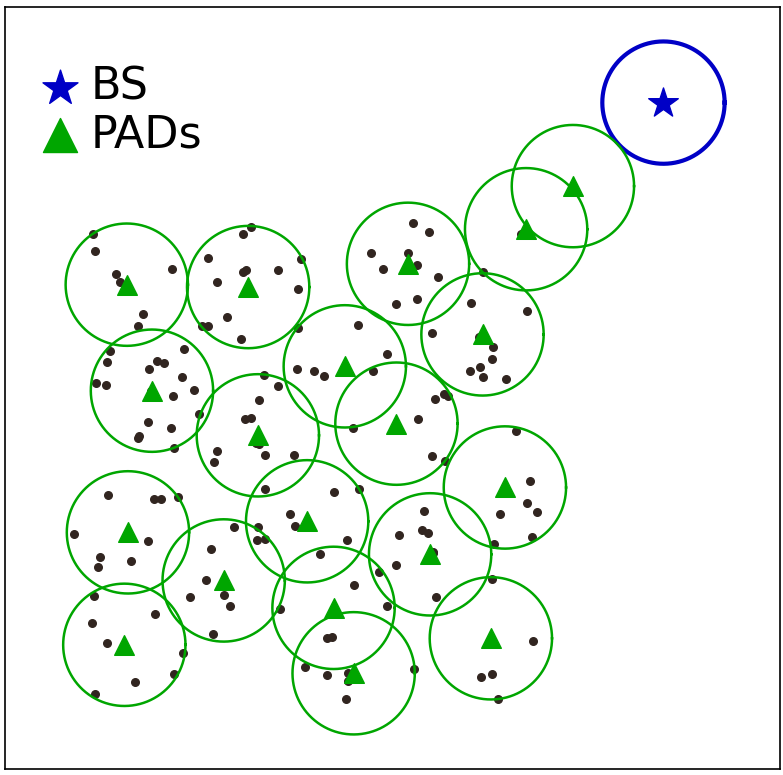}
		\end{minipage}
		\label{fig.2c}
	}
	\centering
	\caption{The result of the DSC algorithm based on Fig.~\ref{fig.1}. From \ref{fig.2a} to \ref{fig.2c}, we show the process to get the minimal PADs.}
	\label{fig.2}
\end{figure*}

The goal of clustering nodes is to determine a series of locations to place PADs to cover the nodes in the network. The node ${s_i} \in C({p_0})$ is surely covered by the BS. Thus, we only cluster the set of nodes $S^{'} = S - C({p_0})$. In our case, we use the K-means algorithm to cluster the nodes. One problem with using the K-means algorithm is that we need to determine the parameter ${K}$. We notice that isolated nodes may appear in the network scenarios since our goal is to be able to handle scenarios with arbitrary node distribution and arbitrary BS locations. The isolated nodes are more difficult to cover than the other nodes. Based on this consideration, we determine the value of $K$ by the number of isolated nodes. Let ${d_i}$ denotes the distance between node ${s_i}$ and its nearest neighbor node ${n_i}$, and ${I}$ is the set of isolate nodes. Besides, we denote $\alpha$ as a parameter to adjust $K$. We have
\begin{equation}\label{eq13}
	I = \{ {s_i}|{n_i} > \sum\limits_{{s_i} \in S'} {{d_i}} /\left| {S'} \right|\}
\end{equation}
\begin{equation}\label{eq14}
	K = \left\lfloor {\alpha \left| I \right|} \right\rfloor
\end{equation}

We use the locations of isolate nodes as the initial points of K-means in Fig.~\ref{fig.1b}. After clustering, we place a PAD in the center of each cluster and generate a coverage disk for each PAD. Then, we check if all nodes are covered. If some nodes are not covered, we select the node farthest from its nearest neighbor node and deploy a PAD in its place $l({s_i})$. We repeat the process until all nodes are covered. For the connectivity constraint, we construct a connected graph from the BS with edge length not exceeding ${d_{max}}$. The PADs closest to the graph are added to the connected graph in turn. If the distance between a PAD to be added and the nearest vertex in the graph exceeds ${d_{max}}$, a new PAD is generated on the line connecting the PAD to the corresponding vertex at a distance ${d_{max}}$ from the vertex, until all PADs are added to the connected graph. The details of the CDC algorithm are shown in algorithm \ref{CDC algorithm}.

\subsection{Disks-shift-combine}\label{DSC}
After obtaining a feasible solution ${P_c}$, and we decrease the number of PADs in the DSC algorithm. To reduce computational complexity, we first delete some redundant PADs in ${P_c}$. The PAD ${p_i}$ is redundant if the ${P_c}$ still satisfies two constraints after removing ${p_i}$. We can observe some redundant PADs in Fig.~\ref{fig.1d} and they are deleted in Fig.~\ref{fig.2a}.

After deleting the redundant PAD, we merge the adjacent PADs to decrease the number of PADs. Before merging the PADs, we first try to reduce the distance between the PADs by a nearest neighbor shift method to facilitate the next merging operation. Let ${Effect}\_C({p_i})$ denotes the set of nodes only covered by the PAD ${p_i}$. We select the PAD with the minimal ${Effect}\_C({p_i}),i \ne 0$ to perform the following operations. We first shift the location of the PAD with the step length ${d_\Delta }$ toward its nearest neighbor PAD, then generate a coverage disk at
the new location, and then check the coverage constraint and the connectivity constraint. If these two constraints are not satisfied, we delete the new location and begin the other loop for the next PAD. Otherwise, we continue to shift this PAD with the step length ${d_\Delta }$. We repeat the above process until all PADs have been shifted. Fig.~\ref{fig.2b} shows the result of shifting. After all the PADs are shifted, we try to re-delete the redundant PADs.

After the shifting operations, we can merge the adjacent PADs to get the minimal set of PADs. We attempt to perform the adjacent PAD merge operation in the following way. For each neighbor PAD ${p_{i}}$ of ${p_{j}}$, $i \ne 0,j \ne 0,i \ne j$, we calculate the discrete point set $C({p_i},{p_j})$, which is the set of nodes only covered by ${p_{i}}$ and ${p_{j}}$:
\begin{equation}\label{eq15}
	C({p_i},{p_j}) = S - \bigcup\limits_{{p_k} \in P,k \ne i,k \ne j} {C({p_k})}
\end{equation}

We select the two most distant nodes in $C({p_i},{p_j})$ and try to merge ${p_{i}}$ and ${p_{j}}$ to a new PAD $p'$ at the midpoint between these two selected nodes. If the merging result can not meet both the coverage constraint and the connectivity constraint, we check whether there is an enclosing circle of radius ${d_{cover}}$ covering $C({p_i},{p_j})$. If there is, we merge ${p_{i}}$ and ${p_{j}}$ to a new PAD $p''$ at the center of this enclosing circle \cite{b16,b17}. We repeat the above steps until all the ${p_i},i \ne 0$ have been tried the merging operation. We show the details in algorithm \ref{DSC algorithm}.

We give a demonstration on how the DSC algorithm works in Fig.~\ref{fig.2}. Fig.~\ref{fig.2a} presents the result of deleting redundant operation on the deployment result in Fig.~\ref{fig.1d}, while Fig.~\ref{fig.2b} and Fig.~\ref{fig.2c} show the results of shifting operations and merging operations, respectively.
\begin{algorithm}  
	\caption{DSC algorithm} 
	\label{DSC algorithm}    
	\hspace*{0.02in}{\bf Input:}
	An initial PAD set ${P}$\\
	\hspace*{0.02in}{\bf Output:} 
	A final PAD set ${P}$  
	
	\begin{algorithmic}[1] 
		\STATE {Check redundant PADs from ${P}$ and delete them}
		\STATE {${P^{'}} = \emptyset$}
		\WHILE [/*Shifting PADs*/]{$P \ne \emptyset$}
		\STATE {Select ${p_i}$ with the minimal $Effect\_C({p_i}),i \ne 0$}
		\WHILE {Two constraints are satisfied}
		\STATE {Shift ${p_i}$ toward its nearest neighbor PAD with ${d_\Delta }$}
		\STATE {$P - \{ {p_i}\} ,{P^{'}} \leftarrow \{ {p_i}\}$}
		\ENDWHILE
		\ENDWHILE
		\STATE {$P = {P^{'}}$, Check redundant again}
		\WHILE [/*Combining PADs*/]{$P \ne \emptyset$}
		\STATE {${N_i} = \{ {p_i}\left|\;{d({p_i},{p_j}) \le {d_{max}},{p_j} \in {P^{'}}} \right.\}$}
		\WHILE {${N_i} \ne \emptyset$}
		\STATE {Select ${s_a},{s_b}$ with the minimal $d({s_a},{s_b})
			$
			\\ ${s_a},{s_b} \in C({p_i},{p_j})$}
		\STATE {Calculate the midpoint ${p^{'}}$ of ${s_a}$ and ${s_b}$}
		\STATE {${P^{''}} = {P^{'}} - \{ {p_i},{p_j}\} ,{P^{''}} \leftarrow {\rm{\{ }}{p^{'}}{\rm{\} }}$}
		\IF {${P^{''}}$ satisfies two constraints}
		\STATE {$P - \{ {p_i}\} ,P - \{ {p_j}\} ,{P^{'}} = {P^{''}}$}
		\STATE {\bf{break}}
		\ELSE 
		\STATE {Select ${s_c}$ with the minimal $d({s_c},{p^{'}})$
			\\ ${s_c} \in C({p_i},{p_j}),c \ne a,b$}
		\STATE { Calculate circumcenter ${p^{''}}$ of triangle with three vertexes ${s_a}$, ${s_b}$ and ${s_c}$}
		\STATE {${P^{''}} - \{ {p^{'}}\} ,{P^{''}} \leftarrow \{ {p^{''}}\}$}
		\IF {${P^{''}}$  satisfies two constraints}
		\STATE {$P \leftarrow {v_{cir}},P - \{ {p_i}\} ,P - \{ {p_j}\} $}
		\STATE {\bf{break}}
		\ENDIF
		\ENDIF
		\STATE {${N_i} - \{ {p_j}\}$}
		\ENDWHILE
		\STATE {$P - \{ {p_i}\}$}
		\ENDWHILE
		\STATE {$P = {P^{'}}$}
		\label{a4}
	\end{algorithmic}  
\end{algorithm}
\section{Simulation Result}\label{simulation}
In this section, we evaluate the performance of the proposed scheme. To illuminate the effectiveness of our proposed scheme, we compare it with the four algorithms in \cite{b13}. To the best of our knowledge, the work in \cite{b13} and our scheme are the only two works investigating the PADs deployment in a UAV-based WRSN.
\begin{figure*}[tp]
	\centering
	\subfigure[]{
		\begin{minipage}[t]{0.46\linewidth}
			\centering
			\includegraphics[width=\textwidth]{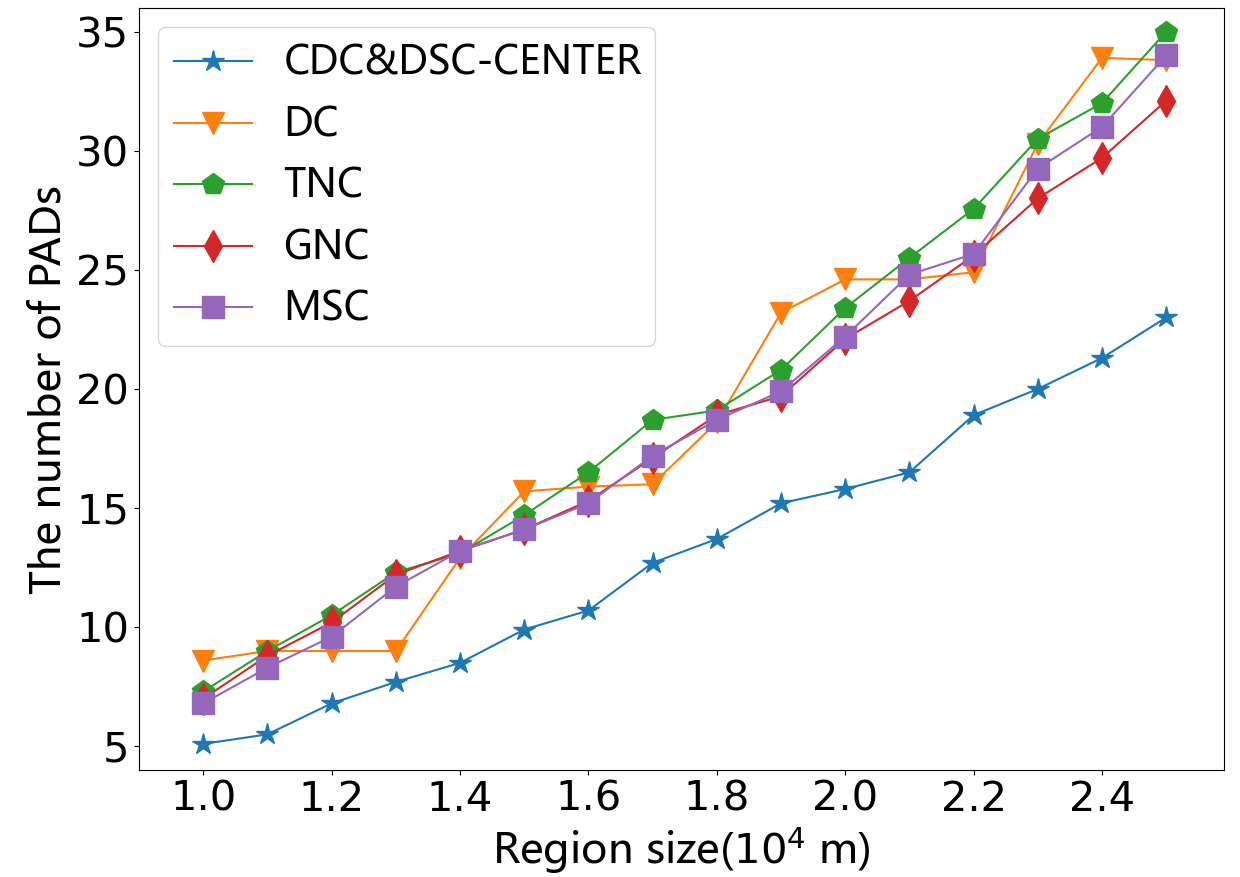}
		\end{minipage}%
		\label{fig.3a}
	}%
	\subfigure[]{
		\begin{minipage}[t]{0.46\linewidth}
			\centering
			\includegraphics[width=\textwidth]{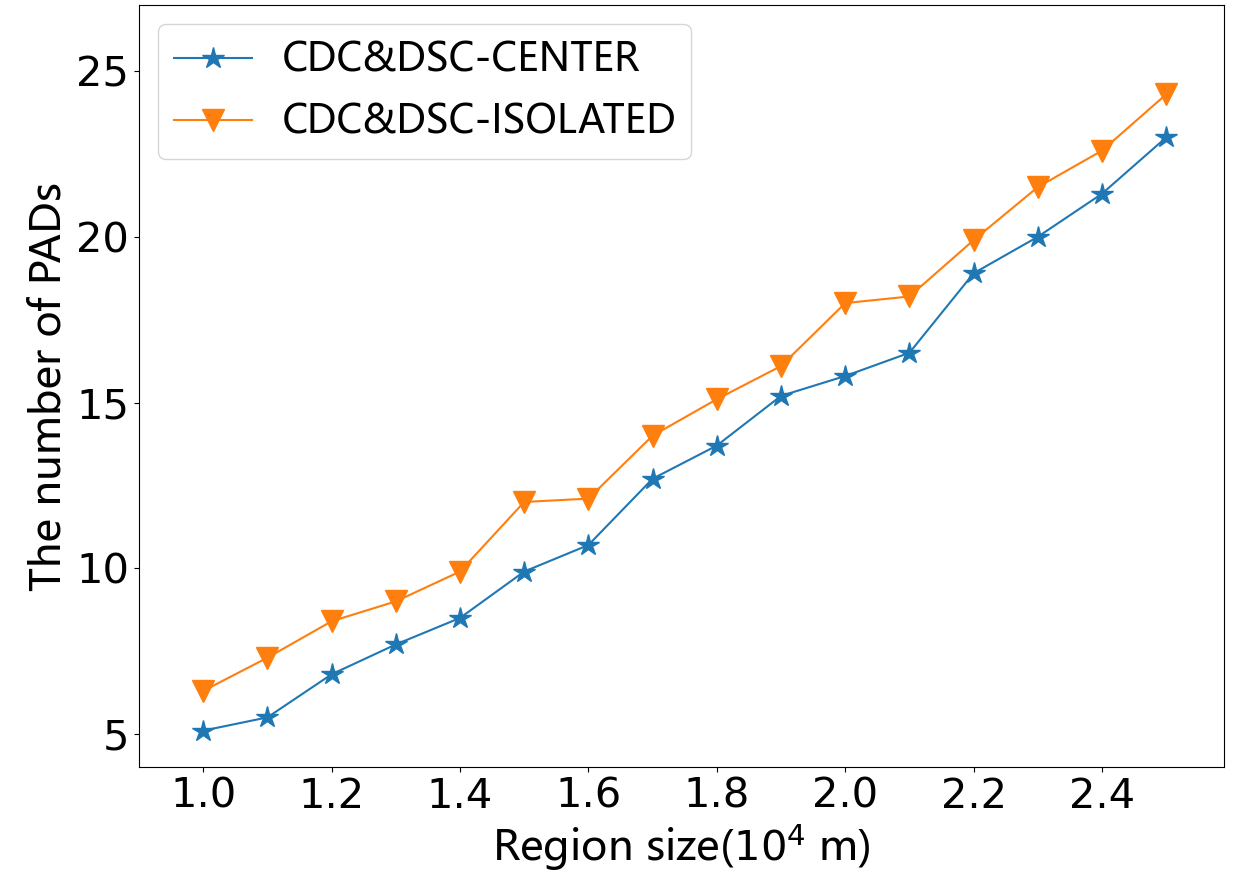}
		\end{minipage}
		\label{fig.3b}
	}
	\centering
	\caption{The result of the region size.}
	\label{fig.3}
\end{figure*}
\begin{figure*}[tp]
	\centering
	\subfigure[]{
		\begin{minipage}[t]{0.46\linewidth}
			\centering
			\includegraphics[width=\textwidth]{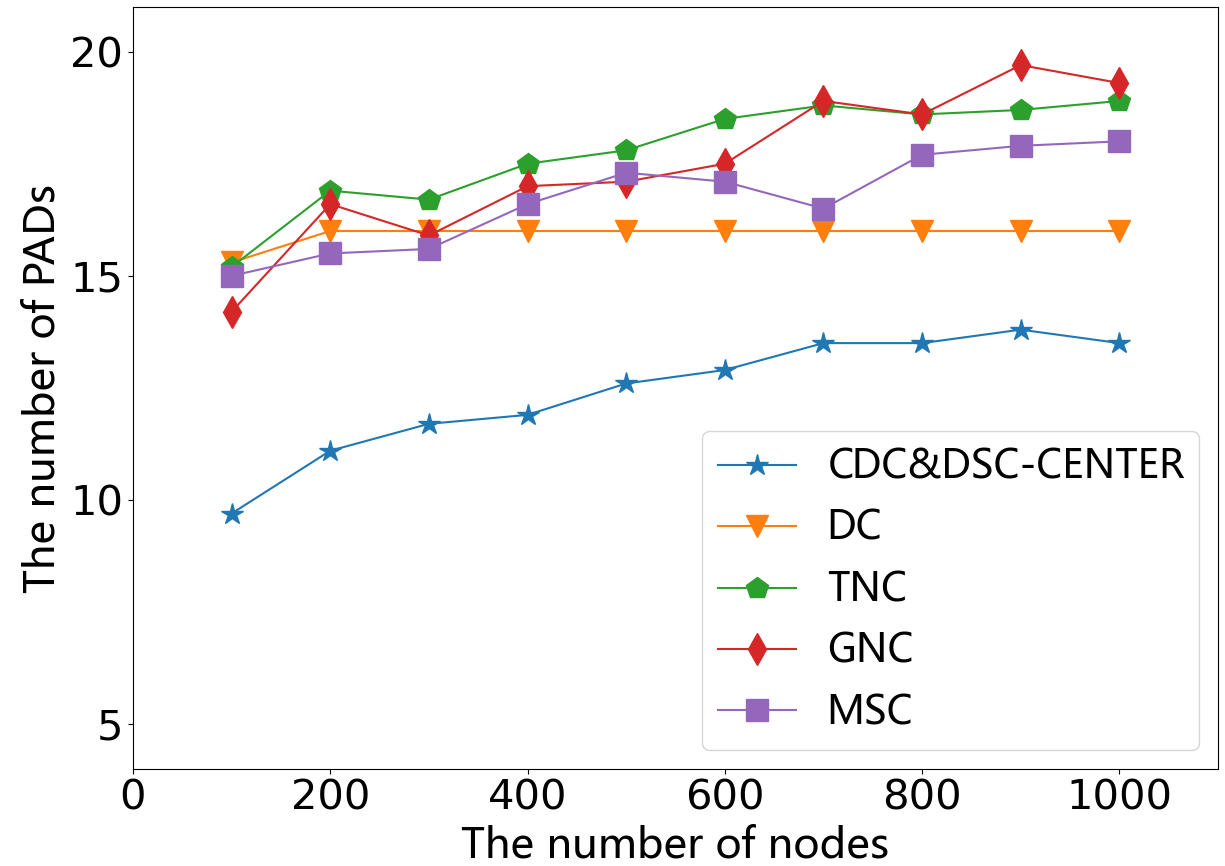}
		\end{minipage}%
		\label{fig.4a}
	}%
	\subfigure[]{
		\begin{minipage}[t]{0.46\linewidth}
			\centering
			\includegraphics[width=\textwidth]{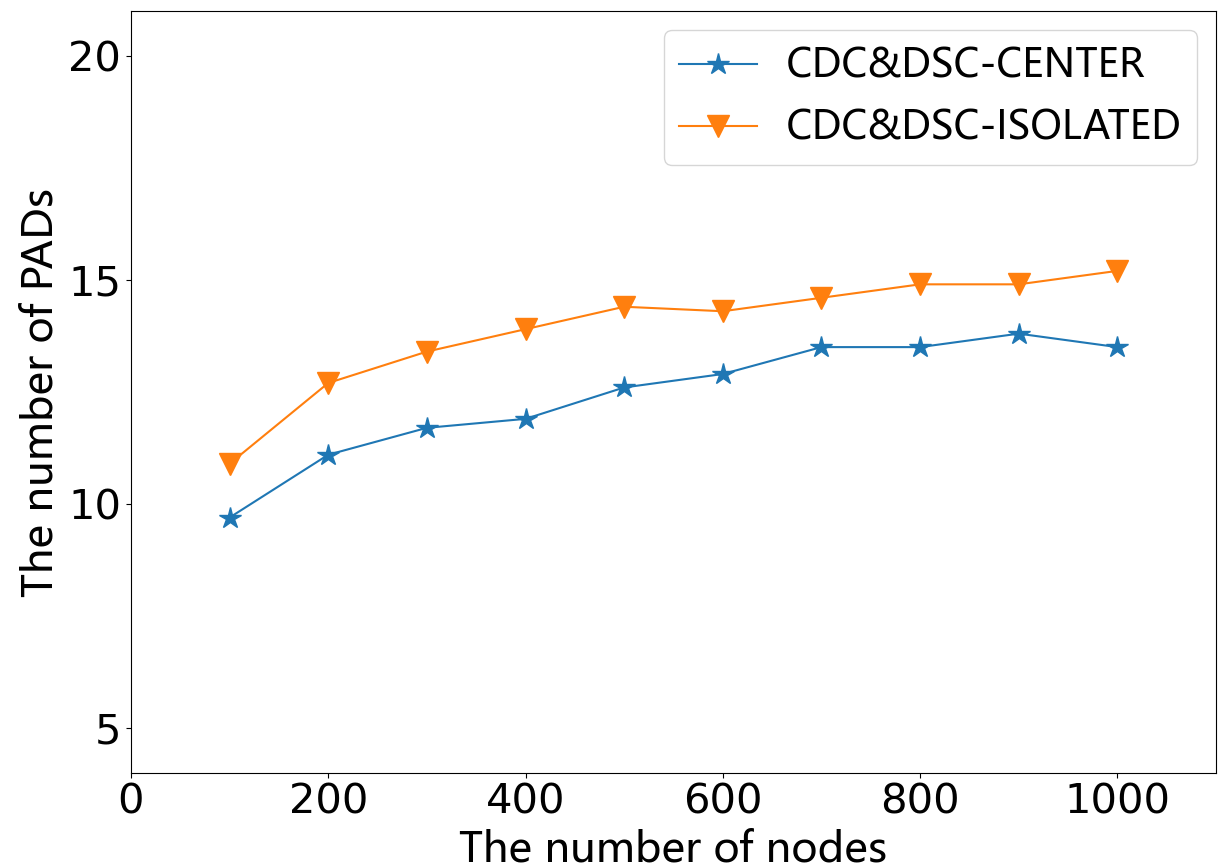}
		\end{minipage}
		\label{fig.4b}
	}
	\centering
	\caption{The result of the number of nodes.}
	\label{fig.4}
\end{figure*} 

We carry out four groups of simulations by varying the size of the network region, the number of nodes, the battery capacity of UAV, and the different node distributions, respectively. To verify the adaptability of the proposed scheme to arbitrary BS locations, we transform two BS locations for each group of simulations: the BS-center scenario, where the BS is surrounded by nodes in the center of the network area, and the BS-isolated scenario, where the BS is outside of the network area and the distance to any node is greater than ${d_{max}}$. Due to the algorithms in \cite{b13} cannot work in the scenarios of BS-isolated, we compare our scheme with the four algorithms in the BS-center scenario first and then we compare the performance of our scheme in these two scenarios. We take the average results with 10 sets of data to smooth data in each simulation. The default parameters are shown in Table \ref{T1}.
\begin{table}[h]
	\small\sf\centering
	\caption{The default parameters. \label{T1}}
	\begin{tabular}{ll}
		\toprule
		Parameters&Value\\
		\midrule
		{Region size(${m}$)}&16000\\
		{Number of nodes}&200\\
		{${E_{max}}({10^4}J)$}&7.8\\
		{${d_\Delta}$}&30\\
		{${\alpha}$}&0.3\\
		{BS location}&[8000,8000](central)\\
		{\quad}&[20000,20000](isolated)\\
		\bottomrule
	\end{tabular}\\[10pt]
\end{table}

\subsection{The impacts of the region size}
Fig.~\ref{fig.3} presents the simulation results with varying the size of the network region from $1000 \times 1000{m^2}$ to $25000 \times 25000{m^2}$. As expected, for all the algorithms, the number of PADs increases as the size of the network region increases. The reason is that as the network size increases, the distance between nodes increases and more PADs are needed to satisfy the coverage constraint. Fig.~\ref{fig.3b} shows the BS-isolated scenario always requires more PADs than the BS-center scenario. This is because we need to introduce additional PADs to ensure the connectivity constraint in the BS-isolated scenario. It is important to notice from Fig.~\ref{fig.3a} that the number of PADs deployed by the proposed scheme is always less than four comparison algorithms, and this advantage continues to grow as the network area increases. This changing trend demonstrates the proposed algorithm has advantages in WRSNs with large network region. 

\subsection{The impacts of the number of nodes}
Fig.~\ref{fig.4} depicts the simulation results for varying the number of nodes from $100$ to $1000$. Comparing with Fig.~\ref{fig.3}, for all the algorithms, the number of PADs is growing slowly as the node number increases in Fig.~\ref{fig.4}. It is because the PAD deployment problem is essentially a coverage problem, and the number of PADs depends mainly on the coverage radius of the UAV and the size of the network, while it is less affected by the node density. We also notice that the result of the DC algorithm is always 16 when the node number is over 200. One reason is that the DC algorithm always deploys PADs at the center of each cell after celling the network area and then removes the PADs of the cells with no node. Therefore, when the node density is above the threshold value (all cells have nodes), the result of the DC algorithm will not change. The other approaches determine the locations of PADs from the geographic distributions of nodes, so an increase in node density has a slight effect on their results. Still, we can find that the proposed scheme achieves the best performance in Fig.~\ref{fig.4a}. From Fig.~\ref{fig.4b}, we conclude the similar conclusion within Fig.~\ref{fig.3b}. In the BS-isolated scenario, it takes more PADs to ensure the connectivity constraint than in the BS-center scenario.
\begin{figure*}[tp]
	\centering
	\subfigure[]{
		\begin{minipage}[t]{0.46\linewidth}
			\centering
			\includegraphics[width=\textwidth]{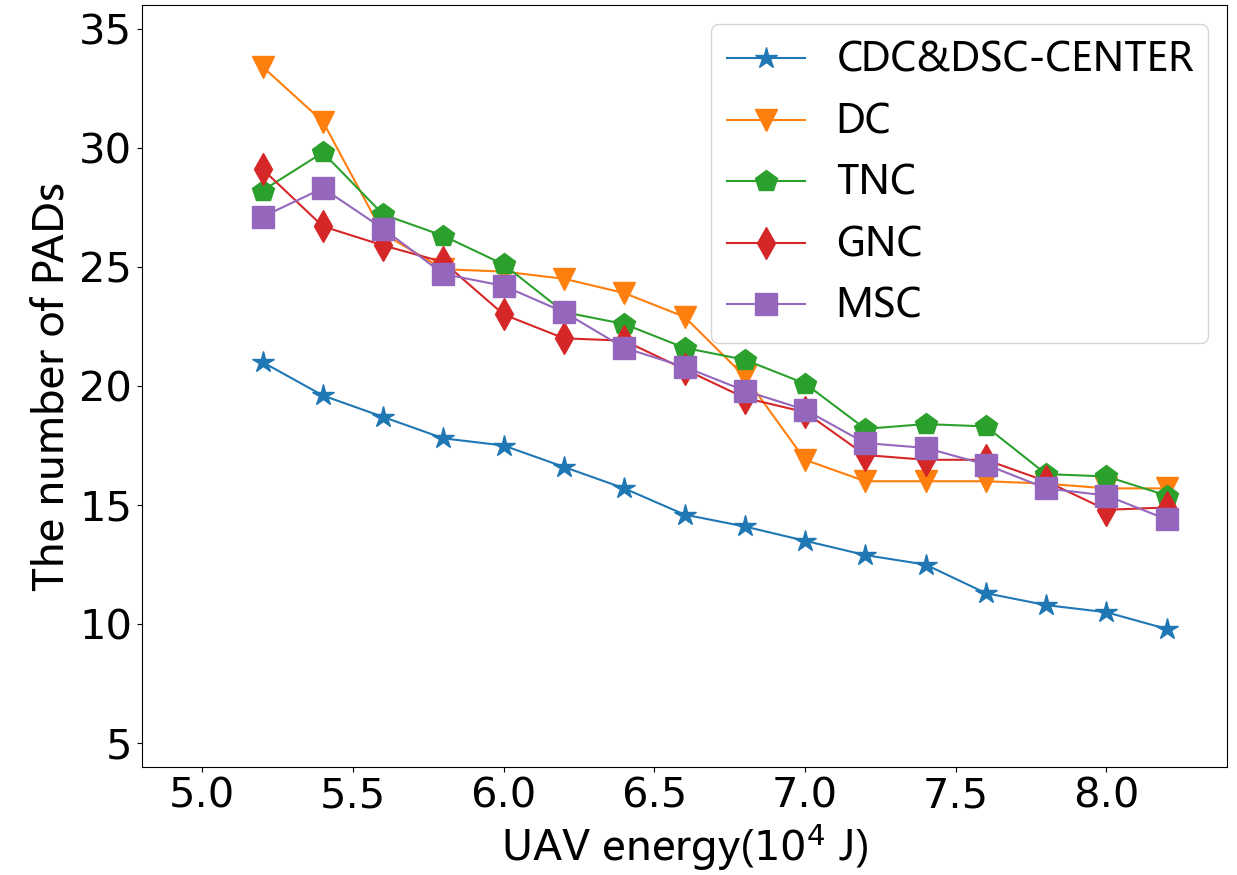}
		\end{minipage}%
		\label{fig.5a}
	}%
	\subfigure[]{
		\begin{minipage}[t]{0.46\linewidth}
			\centering
			\includegraphics[width=\textwidth]{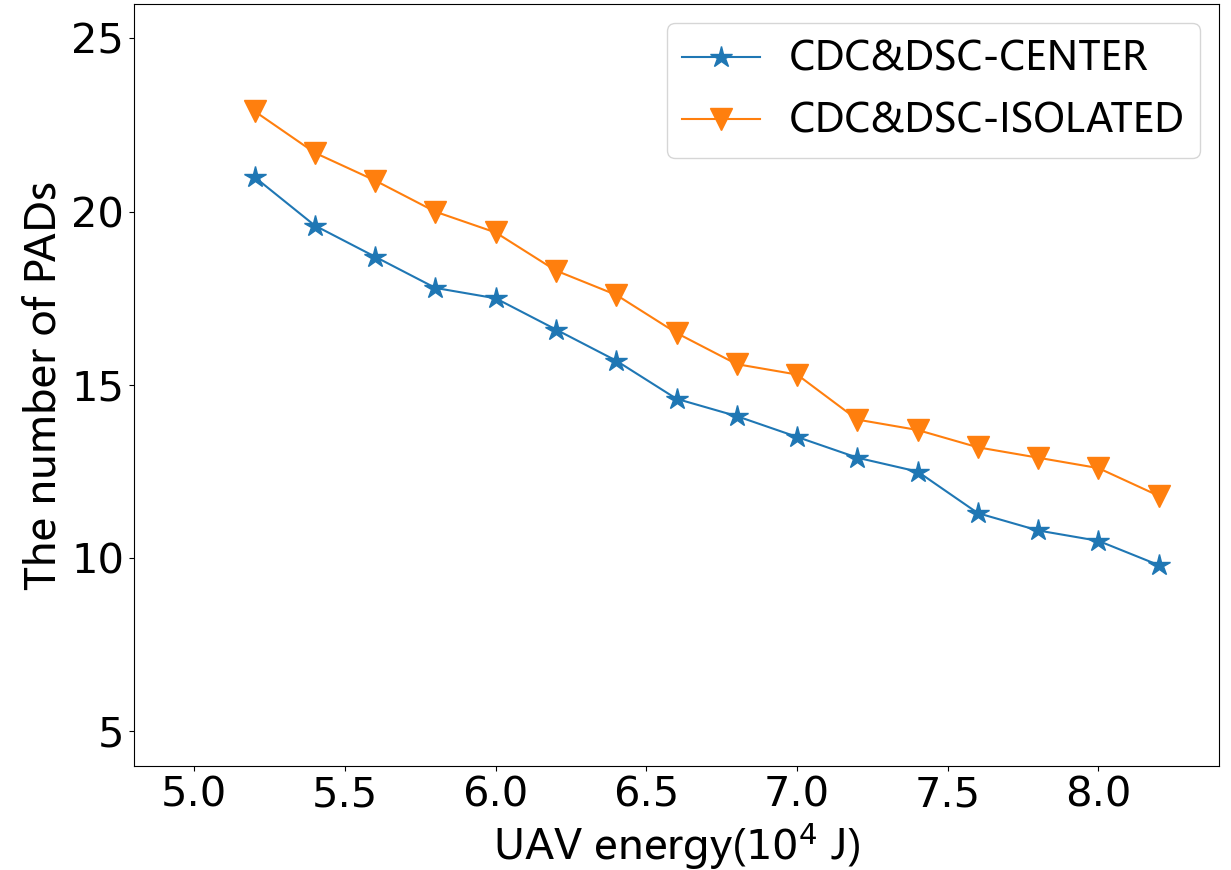}
		\end{minipage}
		\label{fig.5b}
	}
	\centering
	\caption{The result of UAV energy with the uniform nodes.}
	\label{fig.5}
\end{figure*}
\begin{figure*}[tp]
	\centering
	\subfigure[]{
		\begin{minipage}[t]{0.46\linewidth}
			\centering
			\includegraphics[width=\textwidth]{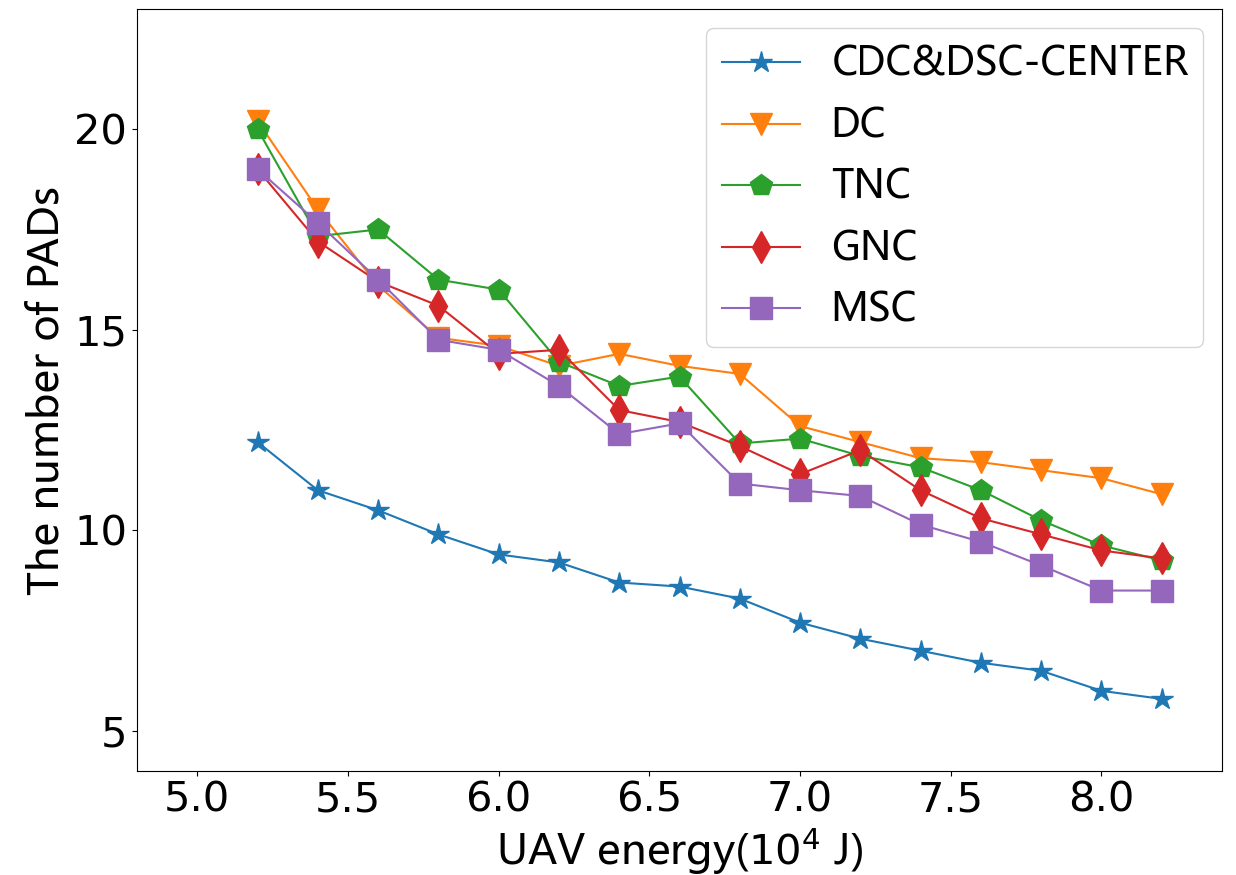}
		\end{minipage}%
		\label{fig.6a}
	}%
	\subfigure[]{
		\begin{minipage}[t]{0.46\linewidth}
			\centering
			\includegraphics[width=\textwidth]{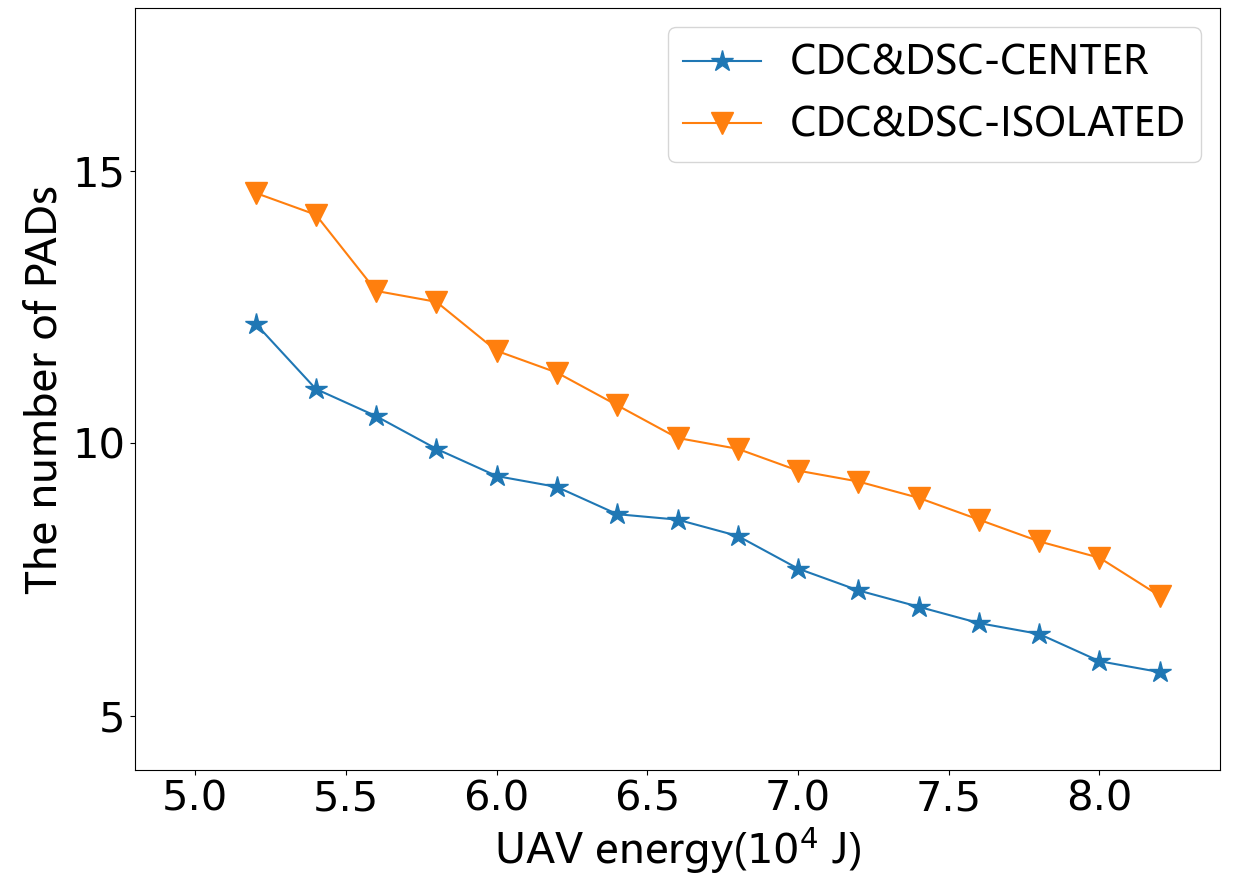}
		\end{minipage}
		\label{fig.6b}
	}
	\centering
	\caption{The result of UAV energy with the mixed Gaussian distribution nodes.}
	\label{fig.6}
\end{figure*}

\subsection{The impacts of the battery capacity of UAV}
We also vary the battery capacity of UAV with results presented in Fig.~\ref{fig.5}. As the battery capacity increases, the number of PADs of all the schemes decreases. The more energy the UAV itself carries, the larger ${d_{max}}$ and ${d_{cover}}$ become. As a result, each PAD has a larger coverage disk, while the connectivity constraint can be satisfied by fewer PADs. The number of PADs in the proposed scheme is obviously 8 to 10 less than all the comparing algorithms. One reason is that the four comparison algorithms always deploy PADs at fixed locations: TNC, GNC, and MSC deploy PADs at nodal locations, while DC deploys PADs at the centers of some cells after celling the network area. Also, none of the four comparison algorithms attempted to merge adjacent PADs.

\subsection{The impacts of the triple Gaussian Mixture distribution}
We further carry out some simulations to verify the adaptability of the proposed scheme to the arbitrary node distribution. We build a scenario with 300 nodes with a triple Gaussian Mixture distribution. We divide 300 nodes into three groups and each group of one hundred nodes is deployed according to a Gaussian distribution. Since the expectation and variance of the three Gaussian distributions are random, the deployment areas of the three groups of nodes may or may not overlap. We vary the battery capacity of UAV in this scenario with results presented in Fig.~\ref{fig.6}. For the DC approach and our proposed scheme, the results are the average of 10 sets of data. Since the TNC, GNC, and MSC may not work in the scenarios with three groups of nodes isolate with each other, for these algorithms we only counted the results in scenarios where they could work. According to Fig.~\ref{fig.6a}, the number of PADs decreases as the battery capacity increases. The reason is the same as the previous simulation results in Fig.~\ref{fig.5a}. Comparing Fig.~\ref{fig.5a}, the numbers of PADs in this type of scenario are much smaller than in the scenarios with a uniform distribution of nodes. It can be explained as follows, when the mixed Gaussian distribution is obeyed, the nodes are clustered more closely than when the uniform distribution is obeyed. Consequently, the actual network area size becomes smaller, the performances of all the approaches are improved in this type of scenario. Our proposed algorithm still achieves the minimum number of PADs in this type of scenario.

In general, the CDC${\&}$DSC algorithm outperforms  all the comparing algorithms. It can adapt to the scenarios with arbitrary node distribution and arbitrary BS location.  

\section{Conclution}\label{Conclution}
After years of research, the introduction of PADs for UAV energy replenishment has become a promising approach to improve the performance of UAV-based WRSNs. In this paper, we investigated the PADs deployment problem for the UAV-based WRSNs. We proposed a novel PADs deployment scheme, named CDC{\&}DSC, to adapt to the scenarios with arbitrary node distribution and arbitrary BS location. We proposed the CDC algorithm to generate an initial deployment of PADs, and then proposed the DSC algorithm to optimize this initial solution in an attempt to merge adjacent PADs and remove redundant PADs by shifting the locations of PADs. Finally, we compared the proposed scheme with four existing PADs deployment approaches through simulations. The results showed that our proposed scheme outperforms the existing methods in all aspects. However, our proposed algorithm deals with the deployment of PADs only from the perspective of minimizing the number of PADs, without considering the charging scheduling. In the future, we are planning to solve the PADs deployment problem to improve the global charging efficiency by taking the charging demand of nodes and the scheduling of UAVs into account.
			
\bibliography{ref}

\begin{thebibliography}{10}
\providecommand{\url}[1]{\texttt{#1}}
\providecommand{\urlprefix}{URL }
\expandafter\ifx\csname urlstyle\endcsname\relax
  \providecommand{\doi}[1]{doi:\discretionary{}{}{}#1}\else
  \providecommand{\doi}{doi:\discretionary{}{}{}\begingroup
  \urlstyle{rm}\Url}\fi
\providecommand{\eprint}[2][]{\url{#2}}

\bibitem{b1}
{Ren}, X., W.~{Liang}, and W.~{Xu}.
\newblock Maximizing charging throughput in rechargeable sensor networks.
\newblock In \emph{2014 23rd International Conference on Computer Communication
  and Networks (ICCCN)}. 2014, pp. 1--8.
\newblock \doi{10.1109/ICCCN.2014.6911792}.

\bibitem{b2}
{Guo}, S., C.~{Wang}, and Y.~{Yang}.
\newblock Joint Mobile Data Gathering and Energy Provisioning in Wireless
  Rechargeable Sensor Networks.
\newblock \emph{IEEE Transactions on Mobile Computing}, Vol.~13, No.~12, 2014,
  pp. 2836--2852.
\newblock \doi{10.1109/TMC.2014.2307332}.

\bibitem{b3}
{Han}, G., J.~{Wu}, H.~{Wang}, M.~{Guizani}, and W.~{Zhang}.
\newblock A Multi-Charger Cooperative Energy Provision Algorithm Based on
  Density Clustering in the Industrial Internet of Things.
\newblock \emph{IEEE Internet of Things Journal}, Vol.~6, No.~5, 2019, pp.
  9165--9174.

\bibitem{b4}
{Lin}, C., C.~{Guo}, H.~{Dai}, L.~{Wang}, and G.~{Wu}.
\newblock Near Optimal Charging Scheduling for 3-D Wireless Rechargeable Sensor
  Networks with Energy Constraints.
\newblock In \emph{2019 IEEE 39th International Conference on Distributed
  Computing Systems (ICDCS)}. 2019, pp. 624--633.
\newblock \doi{10.1109/ICDCS.2019.00068}.

\bibitem{b5}
{Wu}, P., F.~{Xiao}, C.~{Sha}, H.~{Huang}, and L.~{Sun}.
\newblock Trajectory Optimization for UAVs’Efficient Charging in Wireless
  Rechargeable Sensor Networks.
\newblock \emph{IEEE Transactions on Vehicular Technology}, Vol.~69, No.~4,
  2020, pp. 4207--4220.
\newblock \doi{10.1109/TVT.2020.2969220}.

\bibitem{b6}
{Baek}, J., S.~I. {Han}, and Y.~{Han}.
\newblock Optimal UAV Route in Wireless Charging Sensor Networks.
\newblock \emph{IEEE Internet of Things Journal}, Vol.~7, No.~2, 2020, pp.
  1327--1335.
\newblock \doi{10.1109/JIOT.2019.2954530}.

\bibitem{b7}
{Cetinkaya}, O. and G.~V. {Merrett}.
\newblock Efficient Deployment of UAV-powered Sensors for Optimal Coverage and
  Connectivity.
\newblock In \emph{2020 IEEE Wireless Communications and Networking Conference
  (WCNC)}. 2020, pp. 1--6.
\newblock \doi{10.1109/WCNC45663.2020.9120738}.

\bibitem{b8}
{Simic}, M., C.~{Bil}, and V.~{Vojisavljevic}.
\newblock Investigation in Wireless Power Transmission for UAV Charging.
\newblock \emph{Procedia Computer Science}, Vol.~60, No.~1, 2015, pp.
  1846--1855.

\bibitem{b9}
{Choi}, C.~H., H.~J. {Jang}, S.~G. {Lim}, H.~C. {Lim}, S.~H. {Cho}, and
  I.~{Gaponov}.
\newblock Automatic wireless drone charging station creating essential
  environment for continuous drone operation.
\newblock In \emph{2016 International Conference on Control, Automation and
  Information Sciences (ICCAIS)}. 2016, pp. 132--136.
\newblock \doi{10.1109/ICCAIS.2016.7822448}.

\bibitem{b10}
{Costea}, I.~M. and V.~{Pleşca}.
\newblock Automatic battery charging system for electric powered drones.
\newblock In \emph{2018 IEEE 24th International Symposium for Design and
  Technology in Electronic Packaging(SIITME)}. 2018, pp. 377--381.
\newblock \doi{10.1109/SIITME.2018.8599208}.

\bibitem{b11}
{Campi}, T., S.~{Cruciani}, M.~{Feliziani}, and F.~{Maradei}.
\newblock High efficiency and lightweight wireless charging system for drone
  batteries.
\newblock In \emph{2017 AEIT International Annual Conference}. 2017, pp. 1--6.
\newblock \doi{10.23919/AEIT.2017.8240539}.

\bibitem{b12}
{Cai}, C., J.~{Wang}, H.~{Nie}, P.~{Zhang}, Z.~{Lin}, and Y.~G. {Zhou}.
\newblock Effective-Configuration WPT Systems for Drones Charging Area
  Extension Featuring Quasi-Uniform Magnetic Coupling.
\newblock \emph{IEEE Transactions on Transportation Electrification}, Vol.~6,
  No.~3, 2020, pp. 920--934.
\newblock \doi{10.1109/TTE.2020.2995733}.

\bibitem{b13}
{Chen}, J., C.~W. {Yu}, and W.~{Ouyang}.
\newblock Efficient Wireless Charging Pad Deployment in Wireless Rechargeable
  Sensor Networks.
\newblock \emph{IEEE Access}, Vol.~8, 2020, pp. 39056--39077.
\newblock \doi{10.1109/ACCESS.2020.2975635}.

\bibitem{b15}
{Griffin}, B. and C.~{Detweiler}.
\newblock Resonant wireless power transfer to ground sensors from a UAV.
\newblock In \emph{2012 IEEE International Conference on Robotics and
  Automation}. 2012, pp. 2660--2665.
\newblock \doi{10.1109/ICRA.2012.6225205}.

\bibitem{b14}
{Zeng}, Y., J.~{Xu}, and R.~{Zhang}.
\newblock Energy Minimization for Wireless Communication With Rotary-Wing UAV.
\newblock \emph{IEEE Transactions on Wireless Communications}, Vol.~18, No.~4,
  2019, pp. 2329--2345.
\newblock \doi{10.1109/TWC.2019.2902559}.

\bibitem{b16}
{Frank}, N. and N.~{Richard}.
\newblock A fast deterministic smallest enclosing disk approximation algorithm.
\newblock \emph{Information Processing Letters}, Vol.~93, No.~6, 2005, pp.
  263--268.

\bibitem{b17}
{Wan}, P., Y.~{Cheng}, and B.~{Wu}.
\newblock An Algorithm to Optimize Deployment of Charging Base Stations for
  WRSN.
\newblock \emph{EURASIP Journal on Wireless Communications and Networking},
  Vol.~63, 2019, pp. 1--9.

\end{thebibliography}
\bibliographystyle{TRR}
			
\end{document}